\documentclass[aps,prl,twocolumn,showpacs,superscriptaddress,groupedaddress]{revtex4}

\usepackage[utf8x]{inputenc}
\usepackage{amsmath,amssymb,amsfonts,subfigure,braket,color,dsfont}
\usepackage[english]{babel}
\usepackage{graphicx} 
\usepackage{sistyle}
\usepackage{listings}
\usepackage{gensymb} 

\renewcommand{\vec}[1]{\boldsymbol{#1}}

\begin{document}


\title{Precision measurements of electric-field-induced frequency displacements of an ultranarrow optical transition in ions in a solid}
\author{S. Zhang}
\affiliation{LNE-SYRTE, Observatoire de Paris, Universit\' e PSL, CNRS, Sorbonne Universit\' e, 75014 Paris, France}
\author{N. Lu{\v c}i\'c} 
\affiliation{LNE-SYRTE, Observatoire de Paris, Universit\' e PSL, CNRS, Sorbonne Universit\' e, 75014 Paris, France}
\author{N. Galland} 
\affiliation{LNE-SYRTE, Observatoire de Paris, Universit\' e PSL, CNRS, Sorbonne Universit\' e, 75014 Paris, France}
\affiliation{Univ. Grenoble Alpes, CNRS, Grenoble INP and Institut N\' eel, 38000 Grenoble, France}
\author{R. Le Targat} 
\affiliation{LNE-SYRTE, Observatoire de Paris, Universit\' e PSL, CNRS, Sorbonne Universit\' e, 75014 Paris, France}
\author{P. Goldner} 
\affiliation{Chimie ParisTech, Universit\' e PSL, CNRS, Institut de Recherche de Chimie Paris, 75005 Paris, France} 
\author{B. Fang} 
\affiliation{LNE-SYRTE, Observatoire de Paris, Universit\' e PSL, CNRS, Sorbonne Universit\' e, 75014 Paris, France}
\author{S. Seidelin}\email{signe.seidelin@neel.cnrs.fr}
\affiliation{Univ. Grenoble Alpes, CNRS, Grenoble INP and Institut N\' eel, 38000 Grenoble, France}
\author{Y. Le Coq}
\affiliation{LNE-SYRTE, Observatoire de Paris, Universit\' e PSL, CNRS, Sorbonne Universit\' e, 75014 Paris, France}

\date{\today}

\begin{abstract}

We report a series of measurements of the effect of an electric field on the frequency of the ultranarrow linewidth $^7F_0 \rightarrow$  $^5D_0$ optical transition of $\rm Eu^{3+}$ ions in an $\rm Y_2SiO_5$ matrix at cryogenic temperatures. We provide linear Stark coefficients along two dielectric axes and for the two different substitution sites of the $\rm Eu^{3+}$ ions, with an unprecedented accuracy, and an upper limit for the quadratic Stark shift. The measurements, which indicate that the electric field sensitivity is a factor of seven larger for site 1 relative to site 2 for a particular direction of the electric field are of direct interest both in the context of quantum information processing and laser frequency stabilization with rare-earth doped crystals, in which electric fields can be used to engineer experimental protocols by tuning transition frequencies. 

\end{abstract}

\pacs{42.50.Ct.,76.30.Kg}

\maketitle

Optically addressable ions as dopants in solids provide an interesting framework for a wide range of contemporary applications in physics. In particular, rare-earth ions in a crystalline matrix are popular candidates for a steadily growing number of applications due to their outstanding coherence properties~\cite{Yano1991,Equall1994,Thiel2011,Goldner2011,Rancic2018}. Applications include both classical~\cite{Berger2016,Venet2018} and quantum~\cite{Nilson2005,Bussieres2014,Walther2015,Maring2017,Dibos2018} information processing schemes, quantum optical memories~\cite{Zhong2017,Laplane2017,Seri2017}, and laser stabilization techniques~\cite{Julsgaard2007,Thorpe2011,Galland2020_OL}.

In this work we use the 580 nm optical $^7F_0 \rightarrow$  $^5D_0$ transition in $\rm Eu^{3+}$ ions in an $\rm Y_2SiO_5$ host matrix (Eu:YSO). This transition has an ultranarrow linewidth, potentially down to 122 Hz at 1.4 K with 100 G magnetic field~\cite{Equall1994}.  The $\rm Eu^{3+}$ ions can substitute for $\rm Y^{3+}$ ions in two different crystallographic sites, referred to as site 1 and 2, with vacuum wavelengths of 580.04\,nm and 580.21\,nm, respectively, and we here investigate both sites. We have recently demonstrated that the optical transition frequency of the ions can be modulated by applying mechanical stress on the crystalline matrix~\cite{Galland2020,Zhang2020} which makes the system promising for optomechanics experiments in which a mechanical motion couples to the ion frequency~\cite{Molmer2016,Seidelin2019}. An alternative way of controlling the frequency is based on the Stark effect, in which an external electric field interacts with the electric dipole moment involved in the transition~\cite{Macfarlane2007,Li2016}. In the Eu:YSO system, the Stark coefficients for the optical transition have so far been measured with a limited precision, and only in one of the two different substitution sites~\cite{Graf1998,Macfarlane2014}. Here, the technique of spectral hole burning (see later), combined with the use of an ultranarrow linewidth laser, allows us to obtain not only more accurate values for the already measured site, but also provide measurements for the other much less sensitive site, which thus requires a higher precision in order to assess the corresponding Stark coefficient.

\begin{figure}[t]
\centering
\includegraphics[width=80mm]{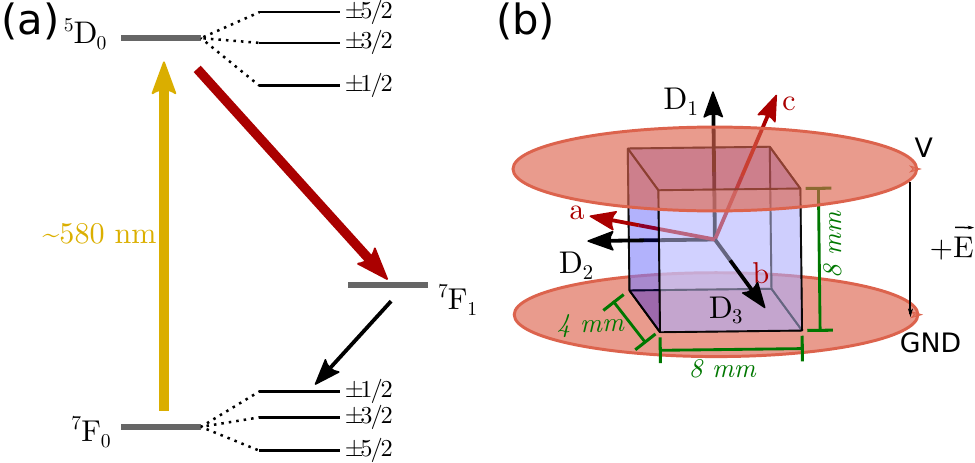}
\caption{\label{levels} Ion level structure and crystal axes. In (a) the energy diagram of the $\rm Eu^{3+}$ ions. After resonantly pumping at 580 nm, the ions decay radiatively primarily to $^7F_1$ and then non-radiatively to one of the hyperfine states which are not resonant with the pump beam, leaving behind a sharp hole in the absorption profile. In (b) we show the  $\rm Y_2SiO_5$ crystal, which is cut parallel to the 3 dielectric axes, D$_1$, D$_2$ and D$_3$ (which is parallel to the crystalline $b$-axis, the $a$ and $c$ axes lying in the D$_1$, D$_2$ plane), as well as the electrodes allowing the application of an electric field $E$ (here, shown parallel to the D$_1$) axis due to an electric potential $V$.}%
\vspace{-0.1cm}
\end{figure}

To study the influence of the electrical field (E-field), we use the technique of spectral hole burning, which allows us to work with a large ensemble of ions without being limited by the inhomogeneous broadening of the collective absorption line, which is of the order of 2 GHz in our case for 0.1-at.\% europium doping. More precisely, the $^7F_0$ ground state consists of 3 hyperfine states (figure~\ref{levels}a), with mutual separations in the 30-100 MHz range~\cite{Yano1991}. Spectral holes are formed by pumping the ions resonantly from the $^7F_0$ to the $^5D_0$ state, from where they decay radiatively, mainly to the $^7F_1$ state~\cite{Konz2003}, before decaying non-radiatively to the three hyperfine states in the $^7F_0$ manifold. Optical pumping prevents population build-up in the hyperfine level with which the pump beam is resonant, creating a transparent window in the inhomogeneous profile at this exact frequency. By subsequently scanning the probe laser in the neighborhood of the pump-laser frequency, the shape and frequency of the spectral hole can be recorded. A spectral hole represents approximately $10^{13}$ $\rm Eu^{3+}$ ions.

Our experimental setup has recently been described in detail elsewhere~\cite{Gobron2017,Galland2020_OL,Galland2020} and will be only briefly summarized in the following. We use two diode lasers (referred to as master and slave) at 1160 nm. The master laser is frequency locked to an ultra-stable Fabry-Perot reference optical cavity by the Pound-Drever-Hall technique, allowing for a frequency instability below $10^{-14}$ for time scales of 1-100\,s and a few Hz laser linewidth. The slave laser is phase locked to the master laser with a tunable frequency offset. The slave laser possesses the same stability as the master, but its frequency is continuously tunable in a range of several GHz. They are both subsequently frequency doubled to reach 580 nm, with an output intensity of approximately 5 mW.  Moreover, an acousto-optic modulator allows us to scan the slave-laser frequency across the spectral structures with a range of approximately 1 MHz.  A high stability (in fact, absolute) optical frequency measurement is provided when necessary by comparing the slave laser against a primary frequency reference via an optical frequency comb. Avalanche photodiodes are used to measure the intensity of the slave laser before and after passing through the crystal, thereby deducing the shape and position of the spectral hole. For maximum absorption~\cite{Konz2003} (both for burning and probing) we propagate the laser beam parallel to the crystallographic $b$-axis, and maintain the laser polarization parallel to the dielectric D$_1$ axis (figure~\ref{levels}b). In this work, the pump and the probe beams are collimated and diaphragmed by an iris of 2 mm diameter (a fourth of the crystal width and height), and kept in the center of the crystal in order to minimize surface effects. The crystal is maintained at 3.2 K in a closed-loop cryostat resting on an active vibration isolation platform. Under these conditions, we obtain spectral holes with a full width at half maximum (FWHM) of about 3 to 5 kHz.

To apply a homogeneous E-field, we use two circular electrodes (31 mm in diameter) above and below the crystal as shown in figure~\ref{levels}b). As the electric permittivity of the crystal is a factor of ten larger than that of vacuum, it is crucial to avoid gaps between electrodes and the crystal. We use a thin ($<$ 0.1 mm) layer of silver lacquer to contact the crystal directly with the electrodes. The distance between the electrodes is consequently the same as the height of the crystal (the electrodes being extended to the surface of the crystal due to the silver lacquer) which is equal to about 8 mm. To study the linear Stark regime, we burn a spectral hole in the absence of an E-field, then apply a voltage ranging from 0 to 28 V, giving rise to an electric field in the range of 0 to 3500 V m$^{-1}$, and observe the influence on the spectral hole.

\begin{figure}[t]
\centering
\includegraphics[width=80mm]{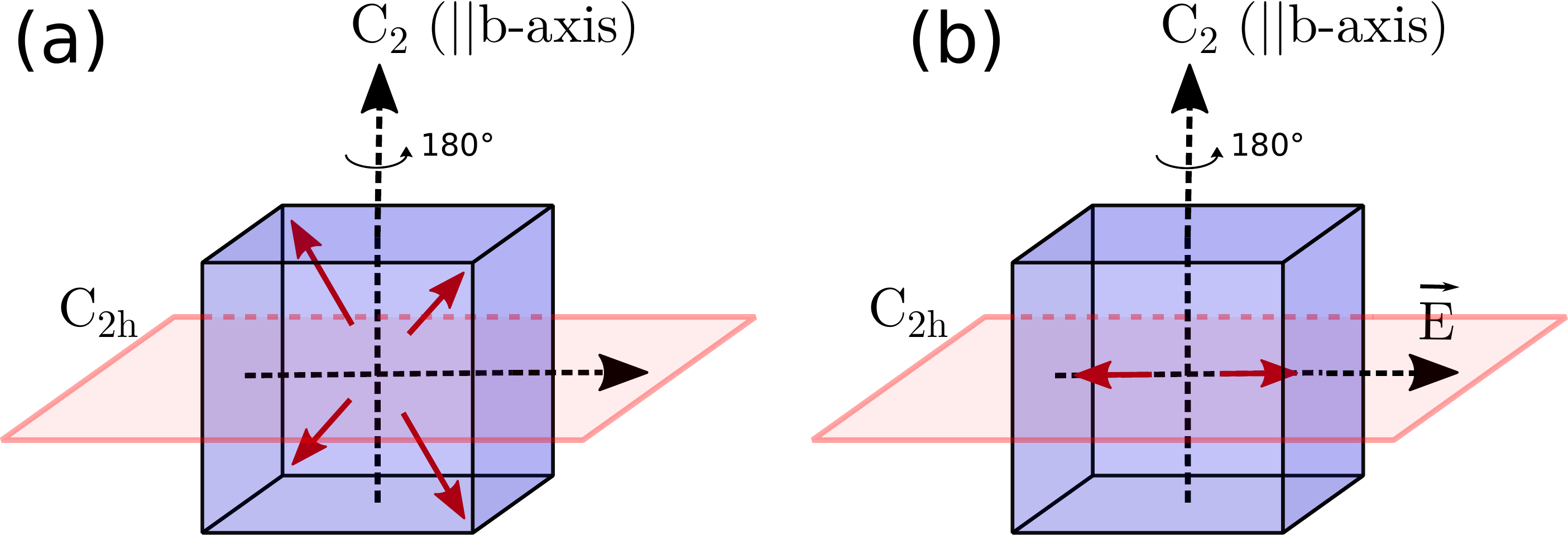}
\caption{\label{symmetry} The symmetry operations of the YSO crystal. In (a) we show the general case, where each $\delta {\vec \mu}_i$ is related by symmetry operations to 4 non-equivalent dipoles. In part (b) we show that when considering the projection of the dipole moment on an axis perpendicular to the C$_2$ rotation axis (thus contained within the horizontal symmetry plane), this number is reduced to 2. Note that the crystal has been rotated by 90$^{\circ}$ relative to figure~\ref{levels} in order to follow the usual convention for which the C$_2$ rotation axis vertical.}
\vspace{-0.1cm}
\end{figure}

The YSO crystal is monoclinic and exhibits a C$_{2h}$ symmetry. The two nonequivalent crystallographic sites  of the $\rm Eu^{3+}$ ions are both of C$_1$ symmetry. In the general case of an application of an E-field with an arbitrary direction, the crystal symmetry gives rise to four possible non-equivalent orientations of the electric dipole moment difference between the ground and excited state ($\delta {\vec \mu}_i$, i=1,2,3,4) for each site, arising from the order of the crystal symmetry group (4) divided by the order of the symmetry group of the crystal site (1), see figure \ref{symmetry}a. Thus, in the general case, the presence of an arbitrary electric field will split a spectral hole into 4 components~\cite{Graf1997}. Because of the centrosymmetric crystal space group, the splitting of the spectral hole will to first order remain symmetric in frequency around zero.

\begin{figure*}[t]
\centering
\includegraphics[width=160mm]{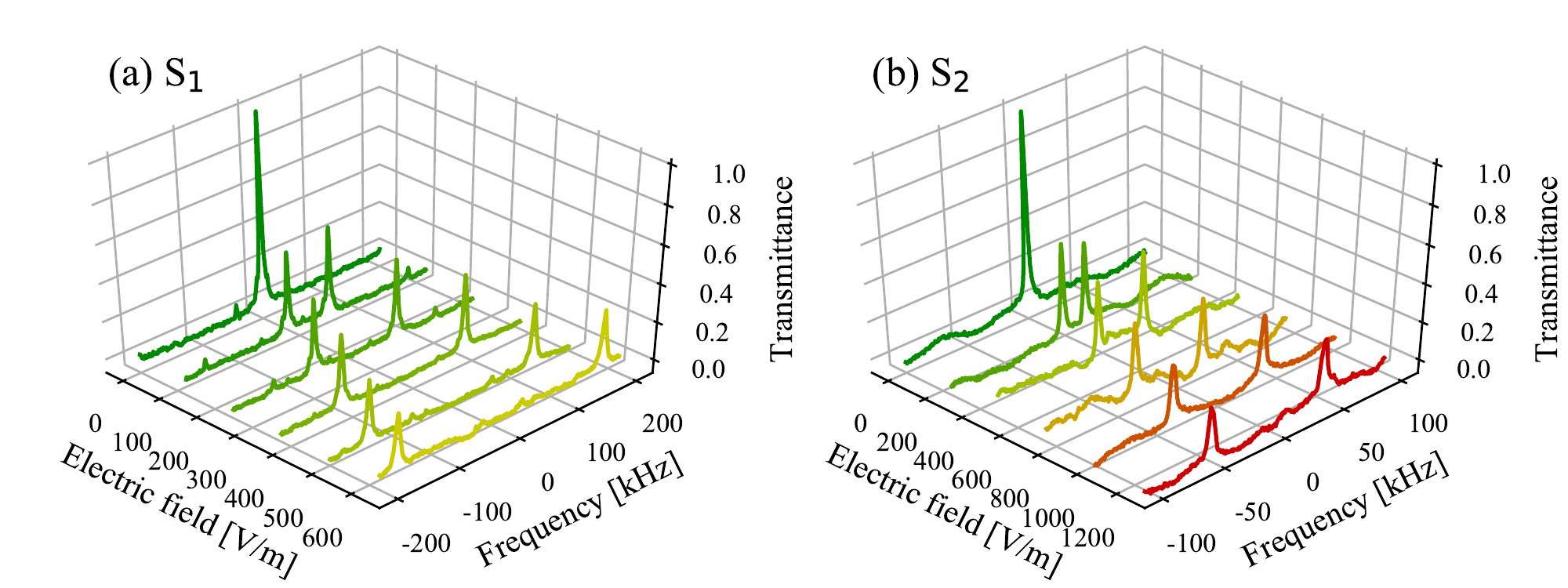}
\caption{\label{3D}Transmittance of an $\rm Eu^{3+}$-doped crystal containing a spectral hole as a function of the electric field applied along the D$_1$-axis for site 1 (a) and site 2 (b). At zero field, the spectral hole exhibits a linewidth below 5 kHz. When an E-field is applied, the spectral structure splits into two components, of which the difference in frequency increases as the E-field is gradually incremented.}
\vspace{-0.1cm}
\end{figure*}

We will determine the Stark effect for an E-field parallel to either the D$_1$ or D$_2$ axis, and thus in both cases, perpendicular to the crystal $b$-axis which coincides with the C$_2$ rotational symmetry axis. As the mirror symmetry plane of the crystal is perpendicular to the C$_2$ rotation axis, the E-field is thus contained within this plane in our case, as illustrated in figure \ref{symmetry}b. This means that the component of a given electric dipole moment perpendicular to mirror plane is not influenced by our E-field, and thus the splitting arising from the mirror symmetry vanishes. Therefore, in all our measurements, the spectral hole only splits into two components, arising from the  C$_2$ rotation symmetry alone.
 
This two-fold splitting into a positive and a negative frequency branch is illustrated for an electric field applied along the D$_1$ direction in figure~\ref{3D}. Here, the transmittance $T$ of the crystal is plotted as a function of frequency difference $f$ with respect to the hole burning frequency, i.e. $f_0\equiv0$, where the initial spectral hole burning occurred, for 6 different values of the applied E-field for the two different crystal sites. The center frequency of each peak is then obtained by a double Lorentzian fit of the transmittance, $T(f) = A + B/\left(1+4(f-f_+)^2/\gamma^2\right) + C/\left(1+4(f-f_-)^2/\gamma^2\right)$, where $f_+$ and $f_-$ denote the center frequency of the two branches, $\gamma$ is the (common) FWHM of the two Lorentzian, and $A$, $B$, and $C$ take into account the offset and normalization of $T(f)$.

\begin{figure}[b]
\centering
\includegraphics[width=85mm]{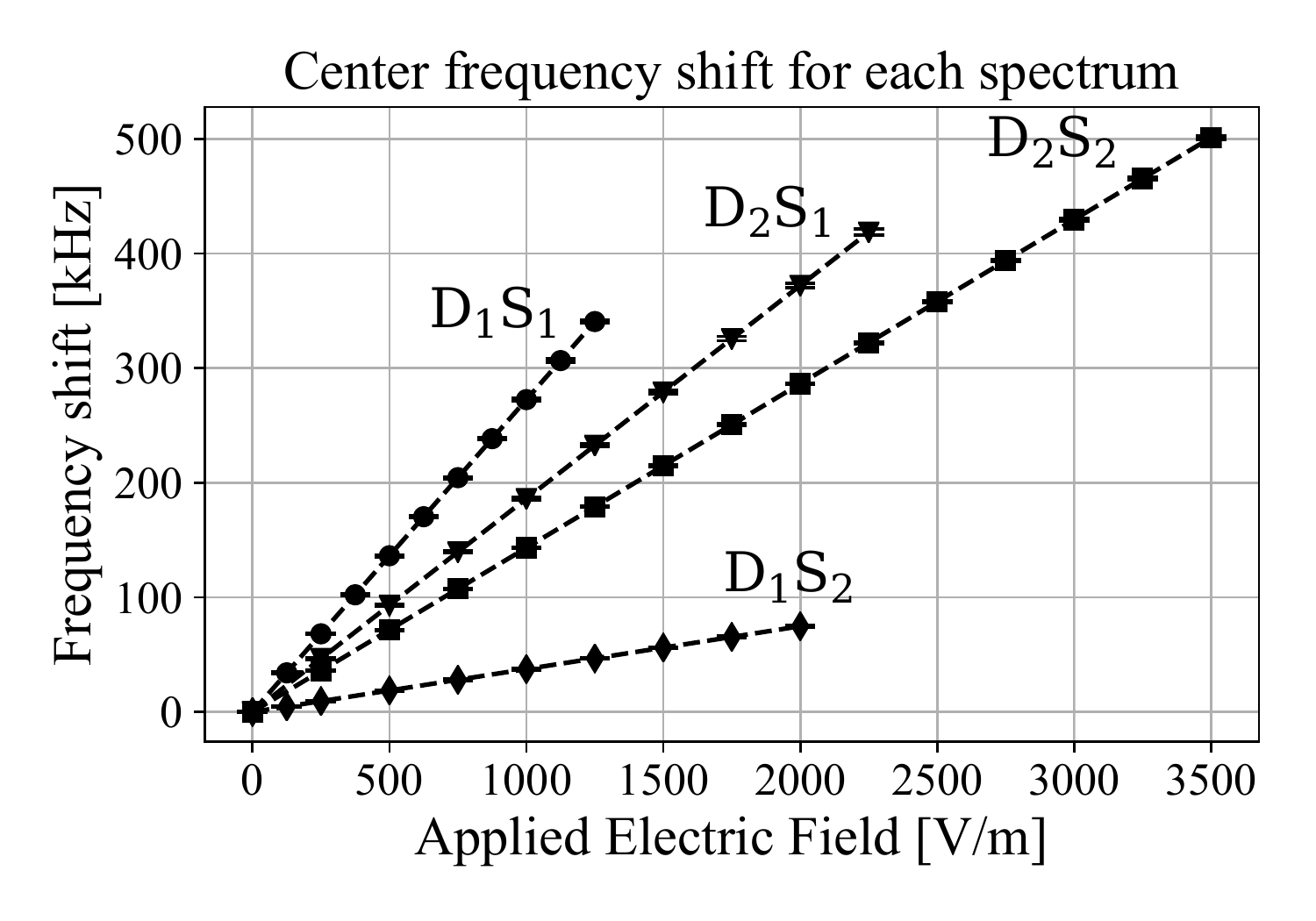}
\caption{\label{linearfits} The displacements of the center frequency of the spectral hole as a function of the amplitude of the E-field applied parallel to D$_1$ and D$_2$, for the two different crystal sites (S$_1$ and S$_2$), as well as linear fits to the data.}
\vspace{-0.1cm}
\end{figure}

These frequency shifts are then plotted as a function of the E-field in figure~\ref{linearfits}, for an E-field aligned along either the D$_1$ or D$_2$ direction, and in each case for both crystal sites. Each point in this figure is the average of two separate measurement sequences or more (a data acquisition sequence lasts approximately 30 minutes). For each sequence, the shift for each data point is half the frequency difference between the positive and negative branch peaks centers, i.e. $\delta f_1 = (f_+-f_-)/2$, (thereby ensuring negligible effect of slow frequency drifts of the reference Fabry-Perot cavity, as such drift would shift both peaks by the same amount in the same direction). The errorbars which are shown on the data and are used in the fit algorithm are estimated individually for each curve, as the noise floor (data baseline) varies from scan to scan and adds to the uncertainty on the determination of the central frequency. They correspond to approximately one tenths of the FWHM of the spectral structures in figure~\ref{3D}. In order to obtain the linear Stark coefficients, we initially assume that the effect is purely linear, and fit the data accordingly with a purely linear model $\delta f_1 = A_1 E$ using a least squares method. The linear Stark coefficients $A_1$ are given in table \ref{tableLinearStark}. Our measurement of the coefficient for site 1 corresponding to an E-field parallel to the D$_1$-axis is consistent with the earlier, less accurate measurement reported in ref.~\cite{Macfarlane2014} (0.27 kHz m V$^{-1}$ with no explicit errorbar). 

\begin{table}
\begin{center}
\begin{tabular}{ccrcc}
		\hline
		Site & Axis & Value & Stat. & Syst.  \\
		\hline\hline
		1 & D$_1$ & 271.8 & $\pm$ 0.5 & $\pm$ 1.6 \\
		1 & D$_2$ & 186.1 & $\pm$ 0.4 & $\mp$ 2.4 \\
		\hline
		2 & D$_1$ & 37.4 & $\pm$ 0.5 & $\pm$ 1.2  \\
		2 & D$_2$ & 143.2 & $\pm$ 0.5 & $\mp$ 0.3 \\
		\hline
\end{tabular}
\caption{\label{tableLinearStark}Linear Stark coefficients for europium ions in a YSO matrix when the electric field is applied parallel to one of the crystalline axes D$_1$ or D$_2$, and for crystal site 1 or 2. These coefficients are given in Hz per (V/m). The $\pm$ and $\mp$ in the last column indicate that the D$_1$ and D$_2$ systematic errors go in opposite directions.}
\end{center}
\vspace{-0.3cm}
\end{table}

The uncertainty on the electric field is composed of three elements: applied voltage, crystal dimensions and inhomogeneity of the electric field arising from an electrode geometrical misalignment and difference in permittivity between vacuum and YSO. The voltage and the crystal dimensions were verified with SI-traceable calibrated instruments, resulting in an error induced by these quantities significantly below the stated uncertainty. We performed numerical simulations with a finite element method software, utilizing the permittivity coefficient of YSO expressed in ref. \cite{Carvalho2015} and allowing for up to 1\,mm misalignment of the electrodes. These simulations confirm that, averaged over the part of the crystal probed by the laser beam, the relative error on the estimation of the electric field is below $3\times10^{-6}$, implying negligible effect on our stated uncertainty. The statistical uncertainty on the coefficients is obtained from the linear fit which takes into account the error-bars on each individual frequency measurement, arising from the lorentzian fits of each spectral pattern. The last column in table \ref{tableLinearStark} shows the systematic errors which arise from the uncertainty on the orientation of the dielectric axis relative to the crystal-cut ($\pm$0.5$^\circ$, as set by X-ray orientation and precision machining). Note that during all measurements, we systematically probe the spectral structure without the E-field before and after the measurement sequence, to verify that no drift has occurred due to other independent physical variables during the measurement sequence.

For applications in which the narrow spectral feature is used for high precision laser frequency stabilization, the linear shifts due to small fluctuations of an electric field only broaden the peak symmetrically (an initial symmetric splitting) and thus does not change the frequency of a laser locked to this peak (only the slope of the frequency discriminator is changed, not its central position). What can potentially be more detrimental in such applications is a higher order Stark effects which would introduce a non-symmetric behavior between the positive and negative branches, and thus would result, for small electric field fluctuations, in laser frequency fluctuations. Due to the high precision in our frequency measurements, we can search for a quadratic component at relatively weak fields, despite the non zero linear sensitivity, and set an upper limit of this type of non-linear Stark shift. We note that this kind of quadratic behavior physically corresponds to a small electric polarizability effect, where E-field-induced additional dipole moments in the ground and excited states, proportional to and along the E-field, produce a correction to the permanent $\delta {\vec \mu}_i$, in a direction (set by the E-field) which is identical for the classes of ions having opposite permanent dipole moments. 

\begin{figure}[t] 
\centering
\includegraphics[width=85mm]{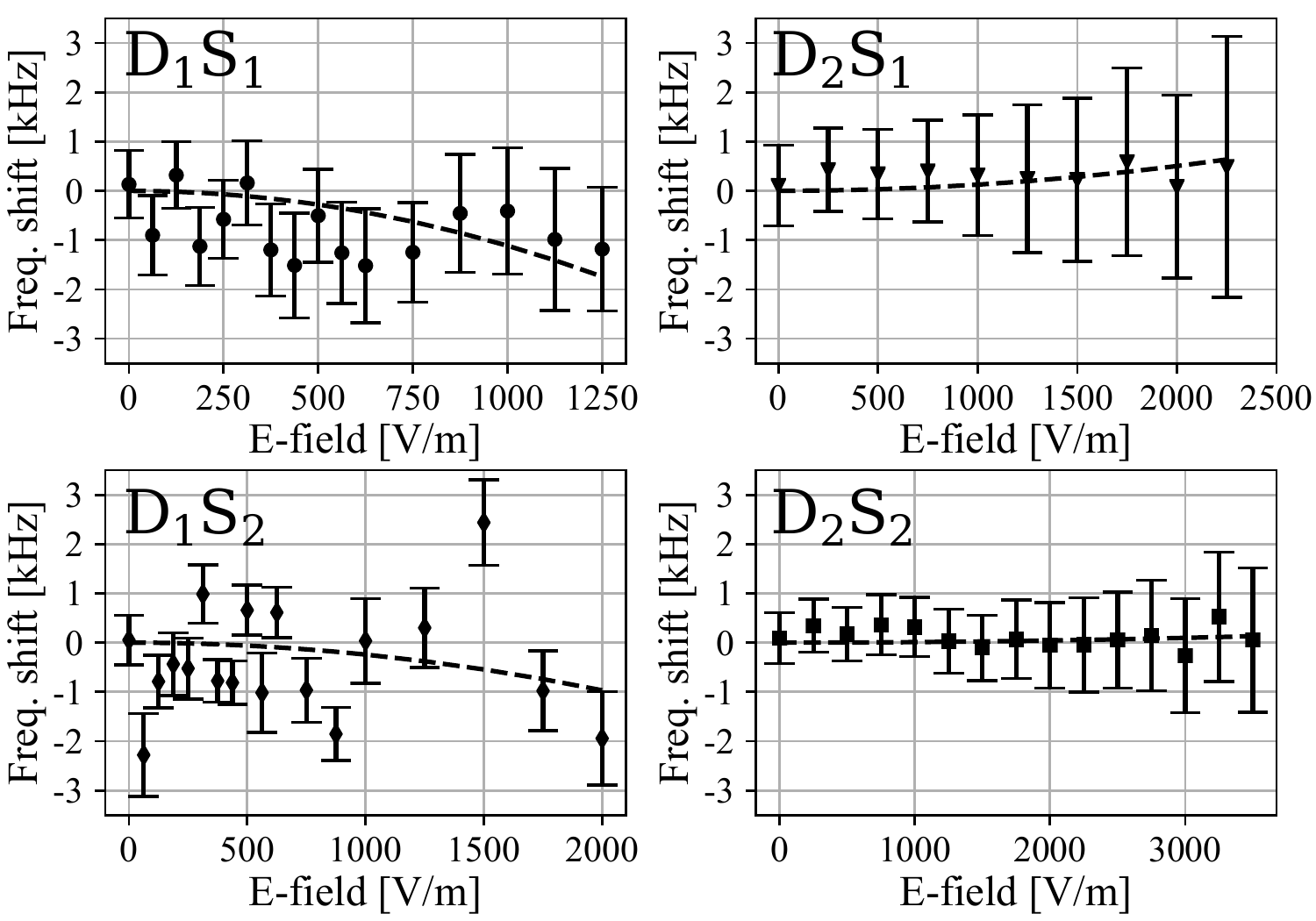}
\caption{\label{Quadratic} Mean central frequency of the positive and negative branch peaks (with reference to the zero E-field case where the initial hole burning occurred) as a function of the applied E-field parallel to D$_1$ or D$_2$, and for site 1 (S$_1$) or site 2 (S$_2$), as indicated by the labels in the insets. The data has been corrected from the drift of the probe laser, which is continuously measured with a primary frequency referenced optical frequency comb. The dashed lines are a quadratic fit to this data, showing an effect compatible with zero to within the measurement uncertainty.}
\vspace{-0.1cm}
\end{figure}

Contrary to the linear (symmetric) case, it is desirable here to explicitly correct for possible slow frequency drift of the reference cavity, typically of the order of $0.1$~Hz.s$^{-1}$, resulting in an apparent drift of frequency of the initial spectral hole, i.e. $f_0=0$ no longer holds throughout the experimental sequence (which lasts longer than stability specification timescale of the cavity). This correction is done by using an optical frequency comb, referenced to primary frequency standard, which is constantly monitoring the probe laser optical frequency. The results are shown, for both directions and sites, in figure~\ref{Quadratic} in which we have plotted $\delta f_2 = (f_++f_-)/2-f_0$ as a function of the applied E-field. We notice the absence of any clear trend corresponding to an asymmetric shift of the peaks in the data. Fitting (least squares method) with a purely quadratic function, $\delta f_2 = A_2 E^2$, we obtain the coefficients $A_2$ for the quadratic Stark effect expressed in table \ref{tableQuadraticStark}. These quadratic coefficient can be considered compatible with zero given the measurement uncertainties, and, as conservative estimates, we set the upper limits stated in the last column of the table. For site 1, this is coherent with the significantly less stringent limit of $\pm$ 0.2 Hz m$^2$ V$^{-2}$ previously obtained~\cite{Graf1998,Thorpe2013}.

\begin{table}
\begin{center}
\begin{tabular}{ccccc}
		\hline
		Site & Axis & Value & Stat. & Abs. val. upper limit  \\
		\hline\hline
		1 & D$_1$ & $-11.1 \times10^{-4}$ & $\pm 5.0 \times10^{-4}$  & $ <3\times10^{-3}$\\
		1 & D$_2$ & $1.3 \times10^{-4}$ & $\pm 2.6 \times10^{-4}$ & $ <1\times 10^{-3}$ \\
		\hline
		2 & D$_1$ & $-2.4 \times10^{-4}$ & $\pm 1.4 \times10^{-4}$ & $ <1\times 10^{-3}$ \\
		2 & D$_2$ & $0.1 \times10^{-4}$ & $\pm 0.5 \times10^{-4}$ & $ <2\times 10^{-4}$\\
		\hline
\end{tabular}
\caption{\label{tableQuadraticStark}Quadratic Stark coefficients for residues after subtracting the linear component. These coefficients are given in Hz per (V/m)$^2$. The systematic error is, here, negligible compared to the statistic error.}
\vspace{-0.3cm}
\end{center}
\end{table}

When the peaks from the positive and negative branches separate under the application of an E-field, a final possibility of quadratic effect corresponds to a deviation from the linear behavior, that would be symmetric between the positive and negative branch. This would physically require the field-induced part of the $\delta {\vec \mu}_i$ to be aligned with the $\delta {\vec \mu}_i$ rather than with the external E-field, but still be proportional to the amplitude of the field. Although this situation is physically more difficult to interpret than the asymmetric quadratic Stark effect discussed above, we still investigate the data for the presence of such behavior. For simplicity, we only consider the positive branch, and examine the residue after subtracting the linear Stark shift determined previously, i.e. $\delta f_{2,+} = f_+-f_0-A_1E$.  These residues are then fitted with a purely quadratic function, $\delta f_{2,+} = A_{2,+}E^2$. From the fitting procedure, we obtain quadratic coefficients for the D$_1$-direction of $(-1.3 \pm 5.1) \times10^{-4} $ Hz m$^2$ V$^{-2}$ for site 1 and $(-0.7 \pm 1.5) \times10^{-4}$  Hz m$^2$ V$^{-2}$ for site 2, and for the D$_2$-direction $(-0.5 \pm 2.6) \times10^{-4} $ Hz m$^2$ V$^{-2}$ for site 1 and $(-0.1 \pm 0.5) \times10^{-4}$  Hz m$^2$ V$^{-2}$ for site 2 (with negligible systematic uncertainties). Again, these coefficients are compatible with zero within the measurement uncertainty and they are all below $\pm$ 1 mHz m$^2$ V$^{-2}$. 

The spectral holes prior to applying an E-field typically exhibits FWHM of approximately 3 to 5 kHz. When the E-field is increased, the spectral structures gradually broaden, as observed in fig.~\ref{3D}. A minor part (approximately 10 \% by the end of 5 consecutive measurements) of this broadening is due to the probing beam, which slowly pumps ions from the edges of the spectral hole into dark states. In principle, the remaining part could be due to a spatial inhomogeneity in the applied field. Finite-element simulations show, however, that such effect is very small ($<$1\% of the linewidth), an assertion corroborated by experimental findings that 1) using a 6 mm diameter probe beam instead of 2 mm does not change the broadening, and 2) highly E-field sensitive cases (eg. D$_1$S$_1$) exhibit similar broadening as less sensitive ones (eg. D$_1$S$_2$). However, a slight misalignemnt of the field (within the stated precision of crystal cutting) producing a component along the $b$-axis may explain most of the broadening (the positive and negative branch broaden symmetrically under such effect). Going to higher E-fields, we have indeed observed a secondary splitting which, extrapolated, accounts for at least 50 \% of the residual broadening. However, no matter the reason behind the broadening, the corresponding uncertainty is largely included in our error-bars on the data, and thus in the uncertainty stated for the Stark coefficients.

In this letter, by means of direct probing of ultranarrow spectral structures, we have provided measurements of the linear Stark effect for an optical transition in $\rm Eu^{3+}$ ions in an $\rm Y_2SiO_5$ matrix by applying a field parallel to the D$_1$ or D$_2$ axis, with an accuracy below the 3 Hz m V$^{-1}$ level. Surprisingly, for an E-field applied parallel to the D$_1$-direction, the  crystal site 1 is a factor of seven more sensitive to the field than site 2. This is useful information for applications using E-fields to electrically tune the transition frequency. Moreover, our measurements have allowed us to provide an upper limit for both an asymmetric and symmetric quadratic Stark coefficient of the order of or below the 1 mHz m$^2$ V$^{-2}$ level, depending on the site, direction and effect considered, an important information for applications in which the electric field is a source of noise, such as metrology applications, in which the transition is used as a frequency reference. For example, in order to reach a relative frequency stability of $10^{-17}$ using a spectral hole as a frequency reference, an asymmetric quadratic Stark effect <1 mHz m$^2$ V$^{-2}$ translates into a requirement of the E-field instability below $2.3$~V m$^{-1}$.\\

The project has been supported by the European Union's Horizon 2020 research and innovation program under grant agreement No 712721 (NanOQTech). It has also received support from the Ville de Paris Emergence Program, the R\'{e}gion Ile de France DIM SIRTEQ, the LABEX Cluster of Excellence FIRST-TF (ANR-10-LABX-48-01) within the Program ``Investissements d'Avenir'' operated by the French National Research Agency (ANR), and the EMPIR 15SIB03 OC18 program co-financed by the Participating States and the European Union's Horizon 2020 research and innovation program. The data that support the findings of this study are available from the corresponding author upon reasonable request.

\end{document}